\documentclass[12pt,preprint]{aastex}

\newcommand{\swf}{{\it Swift}}

\newcommand{\sax}{{\it BeppoSAX}}

\newcommand{\xmm}{{\it XMM-Newton}}

\usepackage{graphicx}
\usepackage{longtable}

\slugcomment{version \today: fm}

\shorttitle{HBLs in the TeV era}
\shortauthors{F. Massaro, A. Paggi, A. Cavaliere 2011}

\begin{document}
\title{X-ray and TeV emissions from High Frequency Peaked BL Lacs}
\author{F. Massaro\altaffilmark{1,2}, A. Paggi\altaffilmark{1}, A. Cavaliere\altaffilmark{3}
}

\affil{Harvard - Smithsonian Astrophysical Observatory, 60 Garden Street, Cambridge, MA 02138}
\affil{SLAC National Laboratory and Kavli Institute for Particle Astrophysics and Cosmology, 2575 Sand Hill Road, Menlo Park, CA 94025}
\affil{Dipartimento di Fisica, Universit\`{a} di Roma Tor Vergata, Via della Ricerca scientifica 1, I-00133 Roma, Italy}

\begin{abstract}
The majority of the extragalactic sources {yet detected at TeV photon energies} belong to the class of ``high frequency peaked BL Lacs" (HBLs) 
that exhibit a spectral energy distribution with a lower peak in the X-ray band.
Such spectra are well described in terms of a log-parabolic shape 
with a considerable curvature, and widely interpreted as synchrotron emission from ultrarelativistic electrons outflowing in a relativistic jet; these are expected to radiate also in \(\gamma\) rays by the inverse Compton process.
Recently we have compared the X-ray spectral parameter distributions of TeV detected HBLs (TBLs)
with those undetected (UBLs), and found that the distributions of the peak energies $E_p$
are \textit{similarly} symmetric around a value of a few keVs for both subclasses, 
while the X-ray spectra are \textit{broader} for TBLs than for UBLs.
Here we propose an acceleration scenario to interpret both the $E_p$ and the spectral curvature
distributions in terms of a coherent and a stochastic acceleration mechanisms, respectively.
We show how the curvature parameter $b\simeq 0.3 - 0.7$ of the synchrotron X rays, that depends only on the latter acceleration
component, can be related to the inverse Compton luminosity in \(\gamma\) rays, so introducing a 
link between the X-ray and the TeV observations of HBLs.
\end{abstract}

\keywords{acceleration of particles - BL Lacertae objects: general - galaxies: active -  radiation mechanisms: non-thermal - X-rays: galaxies}

\section{Introduction}\label{sec:intro}
The BL Lac objects constitute a rare class of Active Galactic Nuclei (AGNs).
Their observational features include: weak or absent emission lines,
high radio and optical polarization, superluminal motions,
and a typical double-humped spectral energy distribution (SED, $\nu F_\nu$).
Their continuum emission is dominated by non-thermal radiations
from radio to $\gamma$-ray frequencies, that make them the most
frequently detected class of extragalactic sources at TeV energies.
The observed broadband emission is widely interpreted as arising in a
jet of relativistic particles closely aligned to our line of sight (l.o.s.) \citep{blandford78}.

In the widely accepted framework 
of leptonic Synchrotron Self-Compton (SSC) radiation,
the low-energy bump is constituted by synchrotron emission
from ultrarelativistic electrons accelerated in the jets; the high-energy 
component is due to inverse-Compton
scattering of these synchrotron photons by the same electron population
\citep[e.g.,][]{marscher85,inoue96}.

The BL Lacs come in two subclasses: the ``low-frequency peaked BL Lacs" 
(LBLs) in which the synchrotron peak
falls in the IR-optical range, and the ``high-frequency peaked BL Lacs" (HBLs) 
where it falls in the UV-X-ray bands \citep{padovani95}.
To mark the HBLs detected at TeV energies from those undetected, 
we refer to the former as TBLs and to the latter as UBLs \citep{massaro11a}.

A convenient description of the BL Lac SEDs has been suggested by Landau et al. (1986) in terms
of a {\it log-parabolic} (LP) model, i.e., a curved, parabolic shape in a double-log plot.
Subsequently, the LP model has been frequently adopted to describe the X-ray spectral continuum
in HBLs \citep{tanihata04,massaro04,massaro08} as well as the TeV 
emission from the TBLs \citep{massaro06,aharonian09}; it has been also used 
\citet{gonzalez2010,giommi11} to describe BL Lac SEDs in the sub-mm and infrared bands.

Such curved spectra are known
to arise both by synchrotron or inverse-Compton radiations
from electron distributions featuring in turn a log-parabolic shape
\citep{massaro04,tramacere07,paggi09}.

Recently, we carried out an extensive investigation of the X-ray synchrotron 
emission of both TBL and UBL subclasses,
based on archival observations carried out by \sax, \xmm~and \swf~ between 
1997 and 2010 \citep{massaro11a}.
On adopting the LP model, the X-ray SED of HBLs is described 
in terms of 3 parameters:
the SED peak position $E_p$, its maximum flux $S_p$ evaluated at 
$E_p$ (or the corresponding peak luminosities
$L_p$ $\simeq$ $4\pi D^2_L S_p$ in terms of the luminosity distance $D_L$), 
and the spectral curvature $b$ around $E_p$
\citep{tramacere07,massaro08}.

Comparing the spectral properties of TBLs and UBLs, we found that:
(i) the $E_p$ distributions of UBLs and TBLs are \textit{similarly} 
symmetric around a value of few keVs for both subclasses; and
(ii) the X-ray spectral curvature $b$ is systematically lower in UBLs than in TBLs,
so that the former feature \textit{narrower} spectral shapes \citep{massaro11a}.

In this Letter we compare the X-ray synchrotron 
luminosities $L_p$ of the TBLs and the UBLs,
as derived from our previous analysis \citep{massaro11a}.
Motivated by the observational results recalled above, 
we interpret both the TBLs and UBLs $E_p$ distributions in terms of a 
coherent electron acceleration scenario,
and those of $b$ as due to accompanying stochastic acceleration.
Finally, we provide a relation between the X-ray spectral curvatures $b$
and the IC luminosities.

We use cgs units throughout this Letter
and assume a flat cosmology with $H_0=72$ km s$^{-1}$ Mpc$^{-1}$,
$\Omega_{M}=0.26$ and $\Omega_{\Lambda}=0.74$ \citep[e.g.,][]{dunkley09}.
In the following, the parameters $E_p$, $S_p$ and $L_p$ refer to the observer reference frame, 
while all unprimed quantities refer to the jet frame.

\section{Log-parabolic Synchrotron Spectra}\label{sec:spectra}
Log-parabolic electron energy distributions (PEDs;
i.e., number of particles per unit  volume and Lorentz factor $\gamma$) are generally written in the form 
$n\left({\gamma}\right)\,=\,n_0\, (\gamma/\gamma_p)^{-2-r\log{\left(\gamma/\gamma_p\right)}}$,
where $\gamma$ is the electron Lorentz factor, $n_0$ the normalization,
$\gamma_p = \langle{\gamma^2}\rangle^{1/2}$ 
the mean particle energy (i.e., the peak of $\gamma^2~n(\gamma)$) and $r$ the electron curvature parameter.
Such PEDs represent the general solution of the energy and time dependent Fokker-Planck 
kinetic equation, that includes systematic (e.g., electrostatic or electrodynamic) 
and stochastic (e.g., turbulent) accelerations, together with radiative and adiabatic cooling as well as 
particle escape and injection terms \citep{kardashev62,paggi09,tramacere11}.

In general, both the peak energy $E_p$ and the peak luminosity $L_p$ of a synchrotron SED 
emitted by a curved electron distribution
depend on $\gamma_{3p}$, the peak energy of the distribution $\gamma^3~n(\gamma)$.
But in the case of a log-parabolic spectrum $\gamma_{3p}$ is proportional to $\gamma_p$ itself to imply
\begin{equation}
\gamma_{3p} = \gamma_p\,10^{1/2r} = \gamma_p\,10^{1/10b}\, ,
\label{eq:gammap}
\end{equation}
given that the relation between the curvature parameter $r$ of the PED and the homologous $b$
of the synchrotron SED reads simply $b =r/5$ \citep[e.g.,][]{massaro06}.
Then for a typical TBL {with} $b \simeq$ 0.3 or UBL {with} $b \simeq$ 0.7 \citep{massaro11a}
the ratio of $\gamma_{3p}$ to $\gamma_p$ is always $\lesssim$ 3.
\begin{figure}[!b]
\includegraphics[height=6.6cm,width=8.5cm,angle=0]{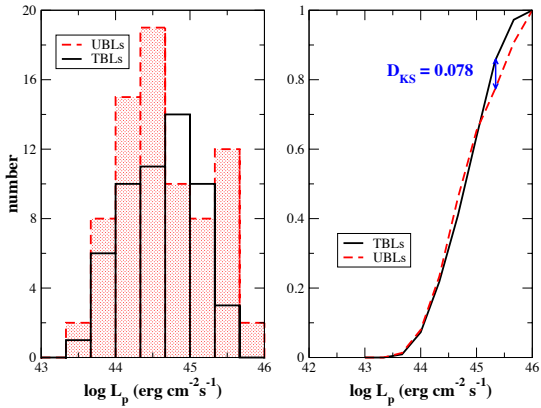}
\caption{The $L_p$ distribution of UBLs (red) and TBLs (black). 
TBL data do not include Mrk 421 and PKS 2155-304 
and the giant flares of Mrk 501 and 1H 1426+421 \citep{massaro11a}.
The maximal separation $D_{KS}$ between the two cumulative distributions, 
(i.e, the KS test variable) is reported.}
\label{fig:histograms}
\end{figure}

The synchrotron peak energy $E_p$ is generally proportional to $\gamma_{3p}^2\,B\,\delta$, where
$B$ is the mean magnetic field and $\delta$ the beaming factor $\delta = \Gamma^{-1}\,(1-\beta cos(\vartheta))^{-1}$,
with $\Gamma$ the bulk Lorentz factor in the jet and $\vartheta$ its opening angle
\citep[e.g.,][]{paggi09,tramacere09}.
In the case of a log-parabolic PED, this relation simplifies to the  
proportionality $E_p\propto\gamma_{p}^2\,B\,\delta\,10^{1/r}$ {on account of Equation~\ref{eq:gammap}} \citep[e.g.,][]{massaro10,massaro11b}.
In addition, the synchrotron peak luminosity $L_p$ scales proportionally to
$n(\gamma_{3p})\,\gamma_{3p}^3\,B^2\,\delta^4$ when considering 
variations of the PED, spectral shape \citep{tramacere11}. But, 
in the case of log-parabolic PEDs, for the product 
$n(\gamma_{3p})\,\gamma_{3p}^3\,=\,N_e\,\gamma_p^2\,f(r)$ obtains, where $N_e$ is the total number of emitting electrons, and
$f(r)$ is a spectral factor depending only on $r$ and ranging from 2.5 to 1.3 
for the typical values of spectral curvature of the TBLs and UBLs, respectively.

We compare the $L_p$ distributions of the two HBL subclasses by
performing a KS test, and find that they are indistinguishable 
at a confidence level of 99\% (see Figure \ref{fig:histograms}).
The $L_p$ distribution for the two HBL populations may still differ within one standard deviation. 
As complementary test, we simulate the two distributions of $log(L_p)$  
with the same number of events (i.e., 55 for TBLs and 76 for UBLs),
adopting the statistical approach described in Massaro et al. (2011a).
We considered two gaussian simulated distributions, 
having the same median ($<log(L_p)>$(UBLs) = 43.80 and $<log(L_p)>$(TBLs) 44.17),
the same variance ($\sigma_{L_p}$(UBLs) = 0.06 and $\sigma_{L_p}$(TBLs) = 0.11) of
the observed distributions and spanning the same range of $L_p$.  
Then, we measured the KS variable, $D_{\rm KS,simul}$, between the two simulated distributions.
We built the distribution of $D_{\rm KS,simul}$ repeating the simulations 8000 times and 
we found that the probability to obtain, randomly, the observed $D_{\rm KS}$ is 86\%.
Thus, UBLs and TBLs have the same  $E_p$ and $L_p$ distributions, 
and differ, significantly, only as for the spectral curvature $b$.

Accordingly, we can assume that $\gamma_p$, $B$, and $\delta$ have close
values in both TBLs and UBLs.
Then, given the proportionality relations for $L_p$ and $E_p$ written above, 
these observational results allow to consider a similar
number $N_e$ of emitting electrons for both HBL subclasses.
Moreover, the impact of limited variations of $L_p$ is softened on $\gamma_p$ and even more on $B$, 
since we expect $B \sim \gamma_p^{1/2}$ (see Section~\ref{sec:systematic}); 
furthermore, the values of $E_p$ are closely the same for the two populations, 
and so $L_p$ turns out to be proportional to $\gamma_p^{3/2}$.

\section{Electron acceleration in BL Lac objects}\label{sec:acceleration}
The characteristic energy \(\gamma_p\) of the PED, simply relates 
to the synchrotron SED peak energy by \(E_p\propto \gamma_p^2 \)
as said in Section~\ref{sec:spectra}, 
is mainly set by the systematic acceleration component. 
On the other hand, the stochastic acceleration mechanism
is responsible for the curvature \(r\) of the PED, related to the spectral width $b$.

In the following, we assume the acceleration mechanisms effective in BL Lac jets
to be a combination of systematic coherent acceleration, 
responsible for the energy peak position of the PED,
and of stochastic acceleration, which accounts for the broadened 
PED around its peak related to the lower spectral curvature.

We consider inverse Compton radiative losses
to be subdominant compared to those by synchrotron emission, consistent with the
observational lack of inverse Compton dominance in the HBL SEDs,
(e.g., PKS 0548-322, \citealt{aharonian10}, 1ES 0806+524, \citealt{acciari08}),
for which any external component appears to be necessary to describe their spectral evolution.

\subsection{Systematic acceleration}\label{sec:systematic}
According to \citet{cavaliere02}, BL Lac jets are likely to be powered by the 
Blandford \& Znajek (BZ, \citeyear{blandford77}) {or {the} \citet{blandford82} {mechanism}} 
(see also \citealt{lovelace76,ghosh1997,krolik1999,livio1999}), i.e., {by} the extraction of rotational {power} 
from a spinning supermassive black hole via the Poynting flux associated
with the adjacent magnetosphere.

As discussed by the above authors, the simple force-free condition 
${\underline E} \cdot {\underline B} = 0 $ governing these magnetospheres
is likely to break down at the jet boundaries, due to considerable electric fields
$E \leq B$ parallel to magnetic field (e.g. \citealt{cavaliere80});
such fields are present in the BZ configuration, especially at the jet boundary.
Alternatively, they may result from magnetic field reconnections 
in current layers at the jet boundary (e.g. \citealt{litvinenko96,litvinenko99});
the related systematic acceleration mechanism is primarily electrostatic 
(see \citealt{massaro11b} for the case of gamma-ray bursts (GRBs).

However, these electric fields will be electrodynamically screened 
out by the embedding plasma at distances that exceed
the screening length $d$ defined by
\begin{equation}\label{eq4}
d = \frac{c}{\omega_p} = \left(\frac{\gamma_p~m_e~c^2}{4\pi~e^2~n}\right)^{1/2} = 5.3 \cdot 10^6 \left(\frac{\gamma_p}{n}\right)^{1/2}\mbox{ cm}\,,
\end{equation}
\citep{cavaliere02}. Here $\omega_p$ is the plasma frequency, $\gamma_p$ is the characteristic electron Lorentz factor,
$m_e$ is the electron mass, $e$ its electric charge, $c$ the speed of light, and $n$ the total electron
(e.g., \citealt{massaro11b}).

Accordingly, the electron energy gained for each acceleration step writes
\begin{equation}\label{eq5}
\gamma_a~m_e~c^2 \simeq e~B~d.
\end{equation}
{Assimilating \(\gamma_a=\gamma_p\),} from Equations \ref{eq4} and 
\ref{eq5} we obtain the expression for the typical Lorentz factor of the accelerated electrons
\begin{equation}\label{gammap}
\gamma_p \simeq \frac{1}{4~\pi~m_e~c^2} \frac{B^2}{n} = 9.8 \cdot 10^4 \frac{B^2}{n}\, .
\end{equation}

The peak energy $E_p$
of the synchrotron emission for an electron of Lorentz factor $ \gamma_p \sim$ 10$^5$ falls 
in the X-ray band, {on} adopting {for} the parameters standard values for HBLs 
$\delta \sim$ 25, $n\sim 1$ cm$^{-3}$ and $B\sim$ 1 G 
\citep[e.g.,][]{massaro06,paggi09,celotti08,acciari08,aharonian09}.

The maximal energy radiated by an electron with Lorentz factor 10$^5$ is $E\sim$ 0.05 TeV.
Then, considering again $\delta\simeq$ 25 we expect the peak of the inverse Compton emission to lie {at} around $E\sim$ 0.5 TeV. 
This is consistent with all TBL observations that have a $\gamma$-ray photon index typically 2 or higher in the TeV energy range,
implying the energy peak of their inverse Compton component to lie below a few TeVs.

To complete the above scenario, the limiting electron Lorentz factor $\gamma_{M}$ attained by an electron corresponds to the condition
where the acceleration compensates the radiative {(mainly synchrotron)} losses. This occurs for
\begin{equation}
\gamma_{M} =  \left(\frac{3\,e}{2\,\pi\,\sigma_T\,B}\right)^{1/2} =1.9 \cdot 10^{7} B^{-1/2}\, ,
\end{equation}
{at values considerably higher than given by Equation \ref{gammap} for our standard values of $B$ and $n$.}

Thus, the maximal energy available to the bulk of the electron population
is of order $\gamma_{M}\,m_e\,c^2$ $\sim$ 1 TeV for $B \sim$ 1G, in agreement with the observed TeV spectral ``tail" of TBLs.
Finally, we note that lower values of the magnetic field coupled with high Doppler factors have also been found to provide good fits 
of the HBL spectra during high luminous states \citep[e.g.][]{finke08}.

\subsection{Geometry and timescales}\label{sec:geometry}
\begin{figure}[!htp]
\includegraphics[height=8.7cm,width=6.2cm,angle=-90]{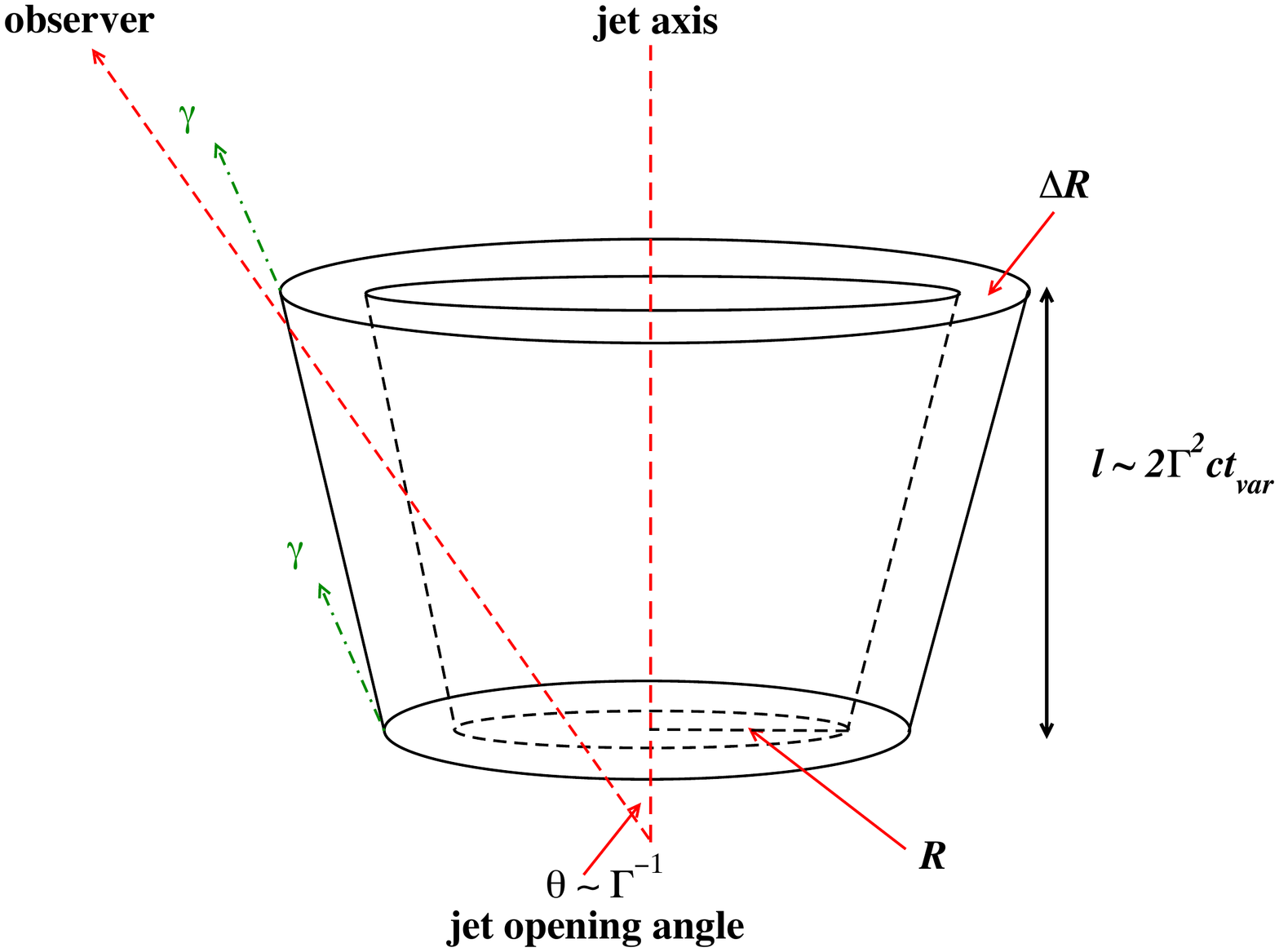}
\caption{A schematic view of the jet emitting region.}
\label{fig:geometry}
\end{figure}

The simplest source condition obtains when the acceleration and the emitting region 
are cospatial.
In particular, we suggest that the emission arises from thin sheaths of thickness $\Delta\,R$
that bound the jet {with radius \(R\gg \Delta R\),} as shown in Figure~\ref{fig:geometry}. There the 
particle density is low, and the
screening length is sufficiently long as to allow the electrons to attain the required high energies
to emit up to TeV energies as discussed above.
 
In this scenario the relativistic aberration of light concentrates radiation isotropically emitted 
in the comoving frame into a cone with opening angle $\vartheta \sim \Gamma^{-1}\ll 1$,
where $\Gamma$ is the bulk Lorentz factor of the jet.
On the other hand, only photons emitted within $\vartheta$ around the l.o.s. will be
detected by the observer. 

The typical delay time $\tau_{d}$ between two photons emitted simultaneously in the
comoving frame from different points on the jet surface is 
\begin{equation}\label{delay}
\tau_{d} = \frac{l}{c}\,[1-\cos(\Gamma^{-1})] \simeq 2\Gamma^2 \frac{l}{c}
\end{equation}
The variability timescale $t_{v}$ consistent with the delay time $t_{d}$
of the photons in the observer's frame implies an upper limit on the physical length
of the emitting region of order
\begin{equation}\label{delay}
l \simeq 2\Gamma^2 t_{v}\,c~.
\end{equation}
This differs from the one usually adopted based on the motion of the jet bulk toward 
the observer and implying $l \simeq \delta t_{v}\,c$. 
Such a ``flashlight" effect is analogous to the one presented by Ryde \& Petrosian (2002) in the case of GRBs, 
but adapted to a cylindrical geometry, and so more like a flash along 
the jet axis over $l \sim 10^{15}$ cm, produced by a relativistic shear instability.

Observed variability timescales $t'_{v}$ of order 10$^3$s imply
$l$ $\sim$ $R$ $\sim$ 10$^{15}$cm to hold for a bulk beaming
factor $\delta\sim$ 25.
This is also consistent with the synchrotron loss length
\begin{equation}
l_{s} \simeq 1.5 \cdot 10^{19} \gamma^{-1}B^{-2} cm\, ,
\label{eq:syn}
\end{equation}
of order 10$^{14}$cm for an electron with 
$\gamma_p\,\sim$10$^5$ in a {field} $B\simeq$ 1 G. 
Considering these emitting regions to lie at the base of the jet close to the supermassive black hole central to the AGN,
the jet radius $R$ can match the height $l$ (i.e., $l\leq R$, see Figure~\ref{fig:geometry}).
Assuming for the thickness of the effective acceleration and emitting region $\Delta R \sim$10$^{-2}\,R$, 
the total volume filled by the emitting electrons is $V \sim \pi/2\,\Delta R\,l\,R \sim$10$^{44}$ cm$^{3}$,
close to the standard values considered for leptonic radiation processes of HBLs 
(e.g., \citealt{massaro06,aharonian09}).

\subsection{Stochastic acceleration}\label{sec:stochastic}
The observational evidence reported in Section \ref{sec:intro} that UBLs feature systematically narrower spectra compared to
TBLs may be interpreted in terms of {a} less efficient stochastic acceleration occurring in the former's jets.

In fact, the curvature parameter of the PED is related to the stochastic acceleration term in 
a Fokker-Planck kinetic equation \citep[e.g.,][]{kardashev62,tramacere07,tramacere09,paggi09,tramacere11},  
and is inversely proportional to the stochastic acceleration rate; 
thus the synchrotron SEDs are relatively broader when the stochastic acceleration is more efficient.
Specifically, the PED curvature $r$
is directly related to the diffusion coefficient $D$ in the Fokker-Planck kinetic equation by
$r \propto D^{-1}$. Higher values of $D$ and faster diffusion also imply
less time spent in the acceleration region \citep[see][]{ginzburg64,protheroe04}.

In a simple statistical picture, $r$ is proportional to the energy gain $\epsilon$ itself,
while it is inversely proportional to the number $n_s$ of acceleration steps, and
to the variance $\sigma^2_{\epsilon}$ of the energy gain; in sum,
$r\propto \epsilon / (\sigma^2_{\epsilon}\,n_s)$ \citep{massaro04,tramacere11}.
As TBLs and UBLs show similar $E_p$ distributions, we can assume that both subclasses 
have similar values of $B$ and of {the} $\epsilon/\sigma^2_{\epsilon}$ ratio.
Thus higher values of $n_s$ correspond to lower values of $r$;
we suggest such high values of $n_s$ to be comparable with smaller acceleration regions,
since each acceleration step is shorter.
So, the observational evidence that $b$ is systematically larger 
in UBLs than in TBLs is consistent with larger volumes $V$ for the former than for the latter. 

Finally, we remark that while the average magnetic field is comparable in TBLs and UBLs, 
the difference is due to the small scale fluctuations in the power spectrum. 
On large scales B is related to the electric field (see Section~\ref{sec:systematic}) 
which is responsible for the systematic acceleration; 
on small scales, a turbulent component gives rise to stochastic diffusion 
(see e.g., Brunetti \& Lazarian 2011, for a related approach concerning radio volumes). 
The latter component 
yields different numbers of acceleration steps in TBLs and UBLs, but averages out on large scales.

On the other hand, considering that the similarities between the $E_p$ and $L_p$ distributions of TBLs and UBLs
have been interpreted in terms of similar
numbers of emitting electrons $N_e$ (see Sect.~\ref{sec:spectra}),
the curvature-volume relation described above suggests that the electron density $n_e\,=\,N_e/V$
is larger in TBLs 
than in UBLs, making a \textit{brighter} inverse Compton peak in a SSC scenario.
We conclude that the TBLs, not only have
\textit{wider} X-ray spectra, but are also expected to be \textit{brighter} in $\gamma$-rays than UBLs.

In addition, the diffusion/acceleration timescale is inversely proportional to the diffusion coefficient; so low values of $D$
will correspond to less variable sources.
This feature is also consistent with the lower variability and the lack of giant X-ray flares found in the whole sample of UBLs
in comparison with the TBLs \citep{massaro11a}.

\section{{Conclusions and} Discussion}\label{sec:discussion}
In \citet{massaro11a}, we analyzed and compared the X-ray spectral properties of TBLs and UBLs,
finding that they have \textit{similar} $E_p$ distributions,
{both} symmetric around a value of few keVs, while
the X-ray spectral curvature $b$ is systematically lower in the former. 

In this Letter, we have compared the $L_p$ distributions of the two HBL subclasses, finding them 
similar at high level of confidence level.
Then UBLs and TBLs differ mainly as for the spectral curvature $b$;
these observational results likely imply similar numbers $N_e$ of emitting 
electrons for both subclasses.

We have proposed to interpret the $E_p$ and $b$ distributions on
assuming that the electron energy gain is due to
both coherent and stochastic particle accelerations.
The scenario is based on re-acceleration rather than 
continuous injection of fresh highly relativistic electrons;
re-acceleration occurs via both systematic and stochastic mechanisms, 
with equilibrium occurring between the overall acceleration rate and the radiative losses.

Describing the coherent acceleration in terms of energy gain from an electric field, we have derived a relation
for the expected particle Lorentz factors $\gamma_p$ $\lesssim$ 10$^5$.
Thus for a typical HBL with magnetic {fields} B $\sim$1 G, plasma density $n\sim$ 1 cm$^{-3}$
and a beaming factor $\delta$ $\sim$ 25, the expected synchrotron peak energy is at $E_p\sim$ 1 keV,
as in fact observed in {the} X-ray SEDs of HBLs \citep{massaro08,massaro11a}.

On the other hand, the stochastic acceleration component is mainly responsible for 
spectral broadening around $E_p$.
In fact, the curvature $b$ of the X-ray spectra is \textit{only} dependent on
the stochastic acceleration term in a Fokker-Planck equation, and thus is inversely proportional to the diffusion coefficient $D$
and to the stochastic acceleration rate $\rho_{acc}$, that is, $b \propto \rho_{acc}^{-1}$. 
Thus, we interpret the narrow X-ray SEDs of UBLs in terms of less efficient stochastic acceleration compared to TBLs.

Finally, pursuing the stochastic acceleration scenario
we have linked the curvature parameter $b$ to the volume of the emitting region, through its
inverse proportionality to the number of acceleration steps $n_s$ (see Section~\ref{sec:stochastic}).
This curvature-volume relation, combined with the above consideration
of similar values of $N_e$, indicates the emitting electron density to be larger in 
TBLs than in UBLs, making the inverse Compton peak \textit{brighter} in the former than in the latter.

Thus electron energies sufficiently \textit{high} to radiate in the TeV range are related to 
sufficient \textit{bright} luminosities for effective detection. Conversely, {narrower}
SEDs and lower fluxes make UBLs harder to 
detect in the TeV range than TBLs, in agreement 
with our previous results concerning their X-ray observations.

\acknowledgements
We thank the referee for the specific suggestions that improved our manuscript.
FM thanks M. Elvis, M. Petrera, J. E. Grindlay, M. Murgia, G. Brunetti and A. Tramacere for fruitful discussions.
FM acknowledges the Fondazione Angelo Della Riccia for the grant awarded him during 2011; 
The work at SAO is supported by the NASA grant NNX10AD50G and by the Foundation BLANCEFLOR Boncompagni-Ludovisi, n'ee Bildt .

~
{}


\begin{thebibliography}{}
\bibitem[Acciari et al. (2008)]{acciari08} Acciari, V. et al. 2008 ApJ 690, 126L
\bibitem[Aharonian et al. (2009)]{aharonian09} Aharonian, F., Akhperjanian, A. G., Anton, G., et al. 2009, A\&A, 502, 749
\bibitem[Aharonian et al. (2010)]{aharonian10} Aharonian, F., Akhperjanian, A. G., Anton, G., et al. 2010 A\&A, 521, 69
\bibitem[Blandford \& Znajek (1977)]{blandford77} Blandford, R. D., Znajek, R. L. 1977 MNRAS, 179, 433
\bibitem[Blandford \& Rees (1978)]{blandford78} Blandford, R. D., Rees, M. J., 1978, PROC. \"Pittsburgh Conference on BL Lac objects",  328
\bibitem[Blandford \& Payne (1982)]{blandford82}Blandford, R.D., \& Payne, D. G. 1982, MNRAS, 199, 883
\bibitem[Cavaliere \& Morrison (1980))]{cavaliere80} Cavaliere, A. \& Morrison, P.  1980 ApJ, 238L, 63
\bibitem[Cavaliere \& D'Elia (2002)]{cavaliere02} Cavaliere, C. \& D'Elia V. 2002 ApJ, 571, 226
\bibitem[Celotti \& Ghisellini (2008)]{celotti08} Celotti, A. \& Ghisellini, G. 2008 MNRAS, 385, 283
\bibitem[Dunkley et al. (2009)]{dunkley09} Dunkley, J., 2009 ApJ, 701, 1804
\bibitem[Finke et al. (2008)]{finke08} Finke, J. D., Dermer, C. D. \& Bottcher, M. 2008 ApJ, 686, 181
\bibitem[Ghosh \& Abramowicz (1997)]{ghosh1997} Ghosh, P., Abramowicz, M. 1997, MNRAS, 292, 887
\bibitem[Giommi et al.(2011)]{giommi11} Giommi, P. et al. 2011 A\&A in press [arXiv:1108.1114]
\bibitem[Ginzburg \& Syrovatskii (1964)]{ginzburg64} Ginzburg, V. L. \& Syrovatskii, S. I. "The origin of Cosmic rays", New York: Macmillan, 1964
\bibitem[Gonz\'{a}lez-Nuevo et al. (2010)]{gonzalez2010}Gonz\'{a}lez-Nuevo, J., de Zotti, G., Andreani, P., Barton, E., J., Bertoldi, F., et al. 2010, A\&A, 518L, 38
\bibitem[Inoue \& Takahara (1996)]{inoue96} Inoue, S., Takahara F., 1996, ApJ, 463, 555
\bibitem[Kardashev (1962)]{kardashev62} Kardashev, N. S., 1962, SvA, 6, 317
\bibitem[Krolik (1999)]{krolik1999} Krolik, J. H. 1999, ApJ, 515, 73
\bibitem[Landau et al. (1986)]{landau86} Landau, R., Golish, B., Jones, T. J., et al. 1986, ApJ, 308, L78
\bibitem[Litvinenko (1996)]{litvinenko96} Litvinenko, Y. E. 1996 ApJ, 462, 997
\bibitem[Litvinenko (1999)]{litvinenko99} Litvinenko, Y. E. 1999 A\&A, 349, 68
\bibitem[Livio (1999)]{livio1999} Livio, M., Ogilvie, G., \& Pringle, J. 1999, ApJ, 512, 100
\bibitem[Lovelace (1976)]{lovelace76} Lovelace, R. V. E. 1976 Natur, 262, 649
\bibitem[Marscher \& Gear (1985)]{marscher85} Marscher, A. P., Gear, W. K. 1985, ApJ, 298, 114
\bibitem[Massaro et al. (2004)]{massaro04} Massaro, E., Perri, M., Giommi, P., et al. 2004, A\&A, 422, 103
\bibitem[Massaro et al. (2006)]{massaro06} Massaro, E., Tramacere, A., Perri, M., Giommi, P., Tosti, G., 2006, A\&A, 448, 861
\bibitem[Massaro et al. (2008)]{massaro08} Massaro, F., Tramacere A., Cavaliere A., et al. A\&A 2008a, 478, 395
\bibitem[Massaro et al. (2010)]{massaro10} Massaro, F., Grindlay, J. E., Paggi, A. 2010a, ApJL, 714, 299
\bibitem[Massaro et al. (2011a)]{massaro11a} Massaro, F., Paggi, A., Elvis, M., Cavaliere, A. 2011 ApJ, 739, 73 (M11)
\bibitem[Massaro et al. (2011b)]{massaro11b} Massaro, F. \& Grindlay, J. E. 2011 ApJ, 727L, 1
\bibitem[Massaro et al. (2011c)]{massaro11c} Massaro, F., Harris, D. E. \& Cheung, C. C. 2011 ApJS in press
\bibitem[Padovani \& Giommi (1995)]{padovani95} Padovani, P., \& Giommi, P., 1995, MNRAS, 277, 1477
\bibitem[Paggi et al. (2009)]{paggi09} Paggi, A., Massaro, F., Vittorini, V. et al. 2009 A\&A, 504, 821
\bibitem[Protheroe \& Clay (2004)]{protheroe04} Protheroe R. J. \& Clay, R. W. 2004 PASA, 21, 1
\bibitem[Ryde \& Petrosian (2002)]{ryde02} Ryde, F. \& Petrosian, V. 2002 ApJ, 578, 290
\bibitem[Tanihata et al. (2004)]{tanihata04} Tanihata, C., Kataoka, J., Takahashi, T., et al. 2004, ApJ, 601, 759
\bibitem[Tramacere et al. (2007)]{tramacere07} Tramacere, A., Massaro, F., Cavaliere, A., 2007, A\&A, 466, 521
\bibitem[Tramacere et al. (2009)]{tramacere09} Tramacere, A., Giommi, P., Perri, M. et al. 2009 A\&A, 501, 879
\bibitem[Tramacere et al. (2011)]{tramacere11} Tramacere, A., Massaro, E., \& A. M. Taylor et al. 2011 ApJ, in press
\end{thebibliography}
\end{document}